\documentclass{SciPost}

\hypersetup{
    colorlinks,
    linkcolor={red!50!black},
    citecolor={blue!50!black},
    urlcolor={blue!80!black}
}

\usepackage[bitstream-charter]{mathdesign}
\usepackage{amsmath}
\usepackage{dsfont}
\usepackage{braket}

\usepackage[normalem]{ulem}

\DeclareSymbolFont{usualmathcal}{OMS}{cmsy}{m}{n}
\DeclareSymbolFontAlphabet{\mathcal}{usualmathcal}

\fancypagestyle{SPstyle}{
\fancyhf{}
\lhead{\colorbox{scipostblue}{\bf \color{white} ~SciPost Physics }}
\rhead{{\bf \color{scipostdeepblue} ~Submission }}

\fancyfoot[C]{\textbf{\thepage}}
}

\begin{document}

\pagestyle{SPstyle}

\begin{center}{\Large \textbf{\color{scipostdeepblue}{
Experimental protocol for observing single quantum many-body scars with transmon qubits\\
}}}\end{center}

\begin{center}\textbf{
Peter Græns Larsen\textsuperscript{1$\star$},
Anne E. B. Nielsen\textsuperscript{1$\dagger$},
André Eckardt\textsuperscript{2$\ddagger$},
Francesco Petiziol\textsuperscript{2$\circ$}
}\end{center}

\begin{center}
{\bf 1} Department of Physics and Astronomy, Aarhus University, DK-8000 Aarhus C, Denmark\\
{\bf 2} Technische Universität Berlin, Institut für Physik und Astronomie, Hardenbergstr. 36, 10623 Berlin, Germany
\\[\baselineskip]
$\star$ \href{mailto:pelar@phys.au.dk}{\small  peter-26-03@live.dk}
$\dagger$ \href{mailto:annebn@phys.au.dk}{\small annebn@phys.au.dk}
$\ddagger$ \href{mailto:eckardt@tu-berlin.de}{\small eckardt@tu-berlin.de}
$\circ$ \href{mailto:f.petiziol@tu-berlin.de}{\small f.petiziol@tu-berlin.de}
\end{center}

\section*{\color{scipostdeepblue}{Abstract}}
\textbf{\boldmath{%
Quantum many-body scars are energy eigenstates which fail to reproduce thermal expectation values of local observables, in systems where the rest of the many-body spectrum fulfils eigenstate thermalization. Experimental observation of quantum many-body scars has so far been limited to models with multiple scar states evenly spaced in energy. It is thus an interesting question whether even \textit{single} isolated scars, which theoretically embody the weakest possibile violation of eigenstate thermalization and may be thought to have no detectable impact in experiments, can leave a trace in measurable quantities. Moreover, single scars offer an interesting scenario for exploring the connection between quantum many-body scars and the original notion of scarring in quantum dynamical systems theory. Here we propose protocols to observe \textit{single} scars in architectures of fixed-frequency, fixed-coupling superconducting qubits. We first adapt known models possessing the desired features into a form particularly suited for the experimental platform. We develop protocols for the implementation of these models, through trotterized sequences of two-qubit cross-resonance interactions, and verify the existence of the approximate scar state in the stroboscopic effective Hamiltonian. Since a single scar cannot be detected from coherent revivals in the dynamics, differently from towers of scar states, we propose and numerically investigate alternative and experimentally-accessible signatures. These include the dynamical response of the scar to local state deformations, to controlled noise, and to the resolution of the Lie-Suzuki-Trotter digitization.
}}

\vspace{\baselineskip}




\vspace{10pt}
\noindent\rule{\textwidth}{1pt}
\tableofcontents
\noindent\rule{\textwidth}{1pt}
\vspace{10pt}


\section{Introduction}
\label{sec:intro}
The discovery of quantum many-body scars (QMBSs) has represented an important twist in the theory of thermalization of isolated quantum systems, with potential applications envisioned in areas like quantum metrology \cite{PhysRevB.107.035123,PRXQuantum.2.020330} and quantum information processing~\cite{PhysRevResearch.4.023095}.

Generic quantum many-body systems are believed to fulfill the eigenstate thermalization hypothesis (ETH)~\cite{DAlessio2016, Gogolin2016}, according to which any eigenstate with finite energy density should yield thermal expectation values of local observables. This behaviour would thus reconcile the reversibility of isolated quantum dynamics with the predictions of equilibrium statistical mechanics. Known exceptions are integrable and many-body localized~\cite{Nandkishore2015, Alet2018, Abanin2019, Sierant2025} systems, with the latter being still under debate~\cite{Abanin2021, Sierant2025}: these systems strongly violate the ETH, in the sense that even all their eigenstates could avoid thermalization, due to the presence (or emergence, in the case of many-body localization) of an extensive number of conservation laws. 
As an intermediate scenario, weak forms of ETH-violation have been identified~\cite{Turner2018,Serbyn2021,Moudgalya_2022}, where only a few eigenstates violate the ETH, in a number which is exponentially smaller than the Hilbert-space size. This is the case of QMBSs, which are thus excited eigenstates that do not give rise to thermal expectation values in otherwise generic quantum many-body systems. 
QMBSs are also related to the more general phenomenon of Hilbert-space fragmentation~\cite{Sala2020}, where the Hilbert space dynamically decomposes into exponentially-many disconnected subspaces with the system size, which are not associated with trivial local symmetries. 

Different theoretical mechanisms have been identified, which lead to the emergence of quantum many-body scarring and which explain the rich landscape of QMBSs that has been discovered. The first phenomenology of QMBSs encountered is as a set of low-entropy eigenstates with equal energy spacing, dubbed towers of states. Towers of QMBSs have been understood to appear, for instance, in models possessing (potentially approximate) so-called spectrum generating algebras~\cite{Moudgalya2020}, where the tower can be constructed by the repeated action of a ladder-type eigenoperator of the Hamiltonian on a suitable eigenstate. The tower of states can in this case be interpreted as the creation of quasiparticles which are non-interacting (due to the equal energy spacing) on top of a given state~\cite{Moudgalya_2022, Chandran2023}. A key feature of towers of QMBSs, which enabled their observation in experiments~\cite{Bernien2017,doi:10.1126/science.abb4928,Zhang2023}, is that they can produce recurrences in the dynamics from certain simple initial states, e.g., in the probability of return to the initial state (Loschimdt echo). This is a consequence of the homogeneous level spacing across the tower, which makes possible for the amplitudes of the initial state on the states of the tower to interfere constructively at certain times. 

Besides models with emergent towers of states, models with an arbitrary number of QMBSs can also be designed by ``manually inserting'' scars at the heart of the spectrum. This can be achieved, for example, by means of so-called projector-embedding methods~\cite{Shiraishi2017, PhysRevB.102.085120, PhysRevResearch.6.L042007}, often combined with matrix-product-state (MPS) formulations~\cite{perezgarcia2007matrix}. This approach highlights that the phenomenon of quantum many-body scarring is not necessarily associated with towers of states, but isolated QMBSs can also exist. Examples include models where individual scars are annihilated by a set of quasilocal operators~\cite{Shiraishi2017, PhysRevResearch.6.L042007}, similarly to the ground states of frustration-free models, as well as frustrated~\cite{Lee2020, Chertkov2021} and topologically ordered~ \cite{Ok2019, Srivatsa2020, PRXQuantum.5.020365} systems.

  While towers of scar states can be diagnosed from the coherent recurrences in the dynamics, as discussed above, individual QMBSs cannot give rise to such an effect and have so far not been observed. The possibility to observe a single scar is, however, an interesting challenge for several reasons. First of all, QMBSs are by definition hard to detect, being excited states immersed in a thermal crowd, whose existence often relies on fine-tuned conditions. A natural question is, then, how weak can the violation of ETH be for actually representing an experimentally relevant and detectable phenomenon: can the ETH be violated even at the single-state level? It has further been shown that individual QMBSs can exhibit novel phenomena, such as ``scar-state phase transitions''~\cite{PhysRevResearch.6.L042007}, where a QMBS embedded in the middle of the spectrum undergoes a phase transition while remaining a QMBS as a parameter is changed.

  Finally, at an even more fundamental level, single QMBSs could be seen as closer 
  counterparts to the original idea of quantum scarring in quantum dynamical systems~\cite{Heller1984, Report}, as opposed to towers of scars. Indeed, from the perspective of quantum chaos theory, the presence of towers of equally-spaced and localized states and the related dynamical recurrences 
  may emerge also from the quantization of locally stable orbits of the corresponding classical system associated with invariant Kolmogorov-Arnold-Moser tori~\cite{Bohigas1993}. It is an interesting observation that for a single quantum many-body scar this scenario can be excluded. The remarkable feature of quantum scarring in its original formulation is, actually, that a weak form of eigenstate localization is possible even in the absence of locally-stable orbits~\cite{Heller1984}. 
 However, such a behavior is generally difficult to prove in quantum many-body systems, which usually do not have a well defined classical analog. Thus, the absence of a tower of states, which might be related simply also to analogs of locally stable KAM-tori, might provide an indication of behavior that is closer to the original idea of quantum scarring. 

In this work, we put forward a protocol to experimentally observe the single scar state predicted in Ref.~\cite{PhysRevResearch.6.L042007}. We first explore candidate models to identify Hamiltonians that are closer to state-of-the-art experimental capabilities. In the models selected, the scar state corresponds to certain known states whose preparation has been already achieved (but not as scars) in experiments, in particular a 1D cluster state~\cite{Raussendorf2021} and an $x$-polarized state. The key interactions required to realize the corresponding parent Hamiltonians are two-qubit Pauli couplings of the form $\sigma_i^z \sigma_j^x$ between neighbouring qubits $i$ and $j$. Interactions of this type appear ``natively" in fixed-frequency superconducting qubits through the so-called cross-resonance effect~\cite{Paraoanu2006, Rigetti2010, krantz2019quantum, Blais2021}. They are indeed employed as the building block of entangling gates in such platforms, such as in IBM Quantum's devices~\cite{qiskit2024}. Given the direct availability of this key element, we propose an experimental realization in an array of fixed-frequency, fixed-coupling transmon qubits. Nonetheless, the protocols developed remain applicable also in other qubit platforms. The different model Hamiltonians considered will then be realized through Lie-Suzuki-Trotter sequences of modified cross-resonance drives.

As discussed above, the experimental observation of single scars cannot rely on coherent recurrences, differently from towers of states, and thus requires novel approaches.
We propose different signatures which take advantage of the accurate single-site control and measurement available in the circuit QED toolbox. In particular, we combine controlled trotterized dynamics with the possibility to measure certain local observables as well as (R\'enyi) entanglement entropies. Moreover, we analyse the impact of (controlled) noise and digitization errors as a probe of the instability of the scar to thermalization. 

The conceived signatures aim at probing general characteristics of a scar state, namely that (1) it is an energy eigenstate, (2) it has low entanglement entropy, and (3) most states near the same energy have entanglement entropies expected for thermal states. Showing these three characteristics experimentally provides convincing evidence of a single scar state. To do so, we propose to prepare the scar state as an initial state and time evolve it with the digital sequence implementing the corresponding parent Hamiltonian for some time. Then, one measures its entropy and the expectation value of a local operator, for which the scar state has some known non-zero value. If, after the time evolution, the expectation value is close to the known value and the entropy is still low, then (1) and (2) are satisfied. If the scar state is an exact eigenstate of the Hamiltonian implemented (more precisely, a Floquet mode of the Lie-Suzuki-Trotter sequence), it will remain stationary during the time evolution. To also show (3), we propose to repeat the time evolution and measurements on a locally deformed version of the scar state. A local deformation will not change the (quasi-)energy much, assuming a local Hamiltonian, but the deformed scar state will have support on many thermal states which are nearby in energy, leading to quick thermalization (high entropy and different expectation value). Comparing the time evolutions of the scar state and its deformed version can thus be used as an experimental signature of the scar state. This method also takes advantage of the fact that scar states are generally efficient to prepare in experiments, unlike generic thermal states.  

In Sec.~\ref{sec:parent Hamiltonians} we briefly review the model of Ref.~\cite{PhysRevResearch.6.L042007} and write down the parent Hamiltonians of the $x$-polarized and cluster states. We propose a blueprint for an experimental implementation of these two parent Hamiltonians in Sec.~\ref{sec:experimental implementation}. In Sec.~\ref{sec:optimal} we numerically test if the scar state can be detected by our proposed method and we check how robust the results are to a random error as well as propose ways to probe the scar's sensitivity. Finally, in Sec.~\ref{sec:conclusion}, we summarize our results and discuss potential further developments.

\section{Parent Hamiltonians}\label{sec:parent Hamiltonians}

In Ref.~\cite{PhysRevResearch.6.L042007}, a family of 1D spin-1/2 models has been proposed which features a single zero-energy scar state $\ket{S}$, based on a matrix-product-state construction. The Hamiltonian is built as the linear combination of $L$ operators ${h}_i$, 
	\begin{gather} 
 \label{eq:sumhi}
		H=\sum_{i=1}^{L}k_i {h}_i,
	\end{gather}
with arbitrary coefficients $k_i\ne 0$. Each operator ${h}_i$ annihilates the scar state, ${h}_i\ket{S}=0$, and acts locally on three consecutive spin sites labelled by $i$, $i+1$, and $i+2$. The general expression of ${h}_i$ depends on two free parameters, $g\in[-1,1]$ and $a\in[0,1]$, and is given in Appendix~\ref{app:A}. In this work, we consider a fixed value $a=1$ and two different values of $g$, namely $g=\pm1$. These choices yield a particularly simple form of ${h}_i$, and thus of $H$, making the model more amenable for experimental implementation. The resulting operators read as
\begin{align}
{h}_i^+ & = - \sigma_i^z + \sigma_i^z\sigma_{i+1}^x, \quad & (\text{for } g=+1) ,\label{eq:x-polarized}\\
{h}_i^- & = - \sigma_i^z - \sigma_{i+1}^x\sigma_{i+2}^z, \quad &(\text{for } g=-1) ,\label{eq:cluster}
\end{align}
where $\sigma_i^\alpha$, with $\alpha\in\{x,y,z\}$, denotes the Pauli matrices. 
Note that the coefficients $k_i$ in Eq.~\eqref{eq:sumhi} need not be invariant under translation by one lattice site. Hence in, e.g., Eq.~\eqref{eq:cluster}, it is important that $\sigma_i^x$ and $\sigma_{i+1}^x\sigma_{i+2}^x$ have the same pre-factor, even though such terms do not overlap spatially.

The scar states associated with ${h}_i^+$ and $h_i^-$ equal certain known states~\cite{PhysRevLett.97.110403}, whose preparation (as non-scars) has been attained in experiments---a feature that will be relevant for our proposed experimental signatures of the scar in Sec.~\ref{sec:optimal}. 
For $g=1$, the scar state is the $x$-polarized state
        \begin{equation}
        \ket{S_+} = \prod_j \ket{+}_j,
        \end{equation}
        where $\ket{+}$ is the $+1$ eigenstate of $\sigma^x$. For $g=-1$, it is the 1D cluster state~\cite{Raussendorf2021}, defined as 
  \begin{equation} \label{eq:cluster_state}
  \ket{S_-} = \left(\prod_{i=1}^{N-1} \mathrm{CZ}_{i,i+1}\right)\ket{S_+},
  \end{equation}
  where $\mathrm{CZ}_{ij}$ is a controlled-$Z$ gate between the $i$th and the $j$th qubit.

\section{Blueprint for experimental implementation}\label{sec:experimental implementation}

We propose an experimental setup for implementing the parent Hamiltonian $H$ of Eq.~\eqref{eq:sumhi} for $h_i=h_i^\pm$ and for probing the physics of a single many-body scar state. The setup consists of a 1D array of fixed-frequency, fixed-coupling transmon qubits~\cite{krantz2019quantum}, as they are implemented, \textit{e.g.}, in the superconducting processors of IBM Quantum~\cite{qiskit2024}. Our first step is to design protocols realizing the necessary qubit-qubit coupling schemes involved in the annihilation operators of Eqs.~\eqref{eq:x-polarized} and~\eqref{eq:cluster}. We will then scale these protocols up via a Lie-Suzuki-Trotter sequence, in order to build the target Hamiltonian of the full system.

\subsection{Building blocks}\label{subsec: cross resonance effect}

The transmon qubits have fixed frequency $\omega_i$ at site $i$ and fixed nearest-neighbor interaction strength $J$. They are far off resonant from each other, $|\omega_i-\omega_j|\gg J$, such that they are effectively uncoupled. The qubit-qubit interactions required in our single-scar models are activated by means of the cross-resonance effect~\cite{Paraoanu2006, Rigetti2010, krantz2019quantum,Malekakhlagh2020}. The latter natively provides the form of interaction needed, which particularly motivates our choice of scarred models and experimental architecture. 

The cross-resonance interaction is obtained by driving the $i$th qubit at the frequency of the $j$th qubit coupled to it. The driven Hamiltonian at the transmon level is described, e.g., in Ref.~\cite{Malekakhlagh2020}. In the two-qubit subspace of the two-transmon Hilbert space, it leads to an evolution captured by an effective Hamiltonian of the form~\cite{krantz2019quantum, Malekakhlagh2020}
	\begin{gather}\label{eq:two site Hamiltonian}
		H_{i,j}=c_{i,j}^{z}\sigma_{i}^z+c_{i,j}^{x}\sigma_{j}^x+c_{i,j}^{zx}\sigma_i^z\sigma_{j}^x+c_{i,j}^{zz}\sigma_i^z\sigma_{j}^z.
	\end{gather}
The coefficients $c_{i,j}^\alpha$, where the first index $i$ labels the control (i.e., the driven) qubit and the second index $j$ labels the target qubit, can be computed in perturbation theory from the driven transmon model~\cite{Malekakhlagh2020} and they are given by 
\begin{subequations}\label{eq:coefficients}
		\begin{gather}
			c_{i,j}^{x}=-\frac{\nu_{j,01}\nu_{i,12}^2}{2(\Delta_{i,j}+\alpha_{i})}J\Omega,
		\end{gather}
		\begin{gather}
			c_{i,j}^{z}=\left(\frac{\nu_{i,12}^2}{4\left(\Delta_{i,j}+\alpha_{i}\right)}-\frac{\nu_{i,01}^2}{2\Delta_{i,j}}\right)\Omega^2,
		\end{gather}
		\begin{gather}
			c_{i,j}^{zx}=\frac{1}{2}\left(\frac{\nu_{j,01}\nu_{i,12}^2}{2\left(\Delta_{i,j}+\alpha_{i}\right)}-\frac{2\nu_{j,01}\nu_{i,12}^2}{\Delta_{i,j}}\right)J\Omega,
		\end{gather}
		\begin{gather}
			c_{i,j}^{zz}=\frac{1}{2}\left(\frac{\nu_{i,01}^2\nu_{j,12}^2}{\Delta_{i,j}-\alpha_{j}}-\frac{\nu_{j,01}^2\nu_{i,12}^2}{\Delta_{i,j}+\alpha_{i}}\right)J^2.
		\end{gather}
	\end{subequations}
Here, $\Delta_i=\omega_{i}-\omega_{j}$ is the detuning between the transition frequencies of two qubits, $\Omega$ is the driving strength and $\alpha_i$ is the level anharmonicity of the $i$th transmon. The coefficients $\nu$ descend from projecting the multi-level transmon Hamiltonian to the two-qubit subspace~\cite{Malekakhlagh2020} and their definition is briefly reviewed in Appendix~\ref{app:B}. Importantly for our purposes, the two-site effective Hamiltonian of Eq.~\eqref{eq:two site Hamiltonian} contains an interaction term $\sigma_i^z\sigma_{j}^x$, which is key for realizing the parent Hamiltonians~\eqref{eq:x-polarized} and~\eqref{eq:cluster}. 

\textit{Case of the $x$-polarized state---} To obtain the fundamental two-qubit building block for the parent Hamiltonian~\eqref{eq:x-polarized} of the $x$-polarized state, the $i$th qubit in the chain is driven at the frequency of the $i+1$st qubit. Then, the cross-resonance Hamiltonian \eqref{eq:two site Hamiltonian} only reproduces the addend in Eq.~\eqref{eq:x-polarized}, if the coefficients in Eq.~\eqref{eq:coefficients} fulfil the following conditions, \begin{subequations}\label{eq:condition}
		\begin{align}\label{eq:condition 1}
		\text{(i)} \quad &	c_{i,i+1}^{z}=-c_{i,i+1}^{zx}, \\
	\label{eq:condition 2}
	\mathrm{(ii)} \quad & c_{i,i+1}^{x}=c_{i,i+1}^{zz}=0.
		\end{align}
	\end{subequations}
These conditions are generally not satisfied directly. We can remedy this by adapting the driving protocol to include additional single-qubit drives that enforce (i) and (ii). Thus, in order to meet the first condition \eqref{eq:condition 1}, a correction of the form $d_{i}^{z}\sigma_i^z$ needs be included in the Hamiltonian, where
 	\begin{gather}
 		d_{i}^{z}=-(c_{i,i+1}^{zx}+c_{i,i+1}^{z}) .
 	\end{gather}
Since $[\sigma_i^z, H_{i,j}]=0$, this can be achieved by applying a (real or virtual~\cite{McKay2017}) single-qubit $z$ rotation, which does not otherwise affect the system. To meet the second condition \eqref{eq:condition 2}, a correction to the $c_{i,i+1}^{x}\sigma_{i+1}^x$ term is needed. This can be achieved by applying a resonant drive to the $i+1$st qubit~\cite{Greenaway2024}, which generates an effective term $d_{i+1}^{x}\sigma_{i+1}^x$ with 
	 \begin{equation} \label{eq:conddi}
	|d_{i+1}^{x}|=\frac{1}{2}\nu_{i+1,01}\Omega_{x,i+1},
   \end{equation}
   where $\Omega_{x,i+1}$ is the driving strength at site $i+1$ and the sign of $d_{i+1}^{x}$ depends on the phase of the drive.
   This term is then calibrated such that $d_{i+1}^{x}=-c_{i,i+1}^{x}$, yielding 
\begin{equation} \label{eq:Omxi}
\Omega_{x,i+1}=\frac{2}{\nu_{i+1,01}}|c_{i,i+1}^{x}| .
	 \end{equation}
  The presence of the resonant drive does not affect the cross-resonance Hamiltonian of Eq.~\eqref{eq:two site Hamiltonian} to leading order in $J$ and $\Omega$, except for the addition of a term $\propto \sigma_{i+1}^x$. 
 To fully meet the condition (ii) in Eq.~\eqref{eq:condition 2}, an additional term compensating the $c_{i,i+1}^{zz}\sigma_i^z\sigma_{i+1}^z$ interaction is also, in principle, needed. Since implementing a two-qubit $zz$ interaction for this purpose would be demanding (being not naturally available in the proposed hardware), we rather render it approximately negligible by an appropriate choice of parameter values. In particular, we will work in the regime of Ref.~\cite{Malekakhlagh2020}, where $J$ is around one order of magnitude smaller than $\Omega$, such that the coefficient $c_{i,i+1}^{zz}$ is also smaller than the others by around one order of magnitude. The presence of a weak but non-zero $c_{i,i+1}^{zz}$ implies that the $x$-polarized state is not an exact eigenstate. Nonetheless, numerical simulations presented in the following confirm that this does not pose relevant limitations towards observing the target scar physics. The final two-site Hamiltonian, with the $x$-polarized state as an approximate zero-energy eigenstate, reads
	\begin{equation}
		H_i^{X}=(c_{i,i+1}^{z}+d_{i}^{z})\sigma_i^z+(c_{i,i+1}^{x}+d_{i+1}^{x})\sigma_{i+1}^x         +c_{i,i+1}^{zx}\sigma_i^z\sigma_{i+1}^x+c_{i,i+1}^{zz}\sigma_i^z\sigma_{i+1}^z.
		 \label{eq:two site Hamiltonian x-polarized}
	\end{equation}
We still display the term $\propto\sigma_{i+1}^x$ , although $c_{i,i+1}^{x}$ and $d_{i+1}^{x}$ are supposed to exactly cancel each other, because we will allow for errors preventing this precise balancing in Sec.~\ref{sec:error}.

\textit{Case of the cluster state---}The cross resonance effect can similarly be used to generate an effective three-site parent Hamiltonian for the cluster state. In this case the $i+1$st qubit is driven at the frequency of the $i$th qubit, generating the Hamiltonian 
	\begin{equation}\label{eq:three site Hamiltonian}
\tilde{H}_i=c_{i+1,i}^{z}\sigma_{i+1}^z+c_{i+1,i}^{x}\sigma_{i}^x+c_{i+1,i}^{zx}\sigma_{i}^x\sigma_{i+1}^z
  +c_{i+1,i}^{zz}\sigma_{i}^z\sigma_{i+1}^z,
	\end{equation}
with the coefficients given in Eqs.~\eqref{eq:coefficients}. Once again, additional single-qubit corrections are needed to reproduce the cluster-state parent Hamiltonian Eq.\ \eqref{eq:cluster}. A resonant drive can be applied to the $i+1$st qubit to generate the term $d_{i}^{x}\sigma_{i}^x$. Equivalently to Eqs.~\eqref{eq:conddi} and \eqref{eq:Omxi}, the compensation condition $d_{i}^x=-c_{i+1,i}^x$ imposes  
\begin{equation}
	\Omega_{x,i}=\frac{2}{\nu_{i,01}}|c_{i+1,i}^{x}|.
\end{equation}
For the cluster state, two different single-qubit terms $d_{i+1}^{z}\sigma_{i+1}^z$ and $d_{i-1}^{z}\sigma_{i-1}^z$ are also needed, with
	\begin{align}
	&	d_{i+1}^{z}=-c_{i+1,i}^{z},\\
	&	d_{i-1}^{z}=c_{i+1,i}^{zx}.
	\end{align}
Combining these three single-qubit modifications with the cross-resonance Hamiltonian \eqref{eq:three site Hamiltonian} yields an approximate three-site building block for the parent Hamiltonian~\eqref{eq:cluster} of the cluster state,
	\begin{equation}
		H_i^{C}=d_{i-1}^{z}\sigma_{i-1}^z+(d_{i+1}^{z}+c_{i+1,i}^{z})\sigma_{i+1}^z  +(d_i^{x}+c_{i+1,i}^{x})\sigma_{i}^x 
 +c_{i+1,i}^{xz}\sigma_{i}^x\sigma_{i+1}^z+c_{i+1,i}^{zz}\sigma_{i}^z\sigma_{i+1}^z. 
  \label{eq:three site Hamiltonian cluster}
	\end{equation}
We next discuss how, equipped with these building blocks, the Hamiltonians for the $x$-polarized and cluster model can be generated on the whole lattice, by means of a Lie-Suzuki-Trotter sequence.

\subsection{Trotterization and stroboscopic Hamiltonian}
\label{sec:eff_ham}

The modified cross-resonance Hamiltonians of Eqs.~\eqref{eq:two site Hamiltonian x-polarized} and \eqref{eq:three site Hamiltonian cluster} are the essential building blocks to obtain approximate parent Hamiltonians of the $x$-polarized and the cluster state, respectively. To get an effective Hamiltonian for a longer chain of transmon qubits, one needs to realize these Hamiltonians on each triplet of neighboring sites. Since the cross-resonance interaction cannot be simultaneously activated for every pair of neighbouring qubits by driving all qubits simultaneously, the protocols will be based on Trotterization (that is, digital quantum simulation or step-wise Floquet engineering). We propose the following protocols for a chain of length $L$ with~$(L~\mathrm{mod}~3)=0$, periodic boundary conditions, and three different transmon frequencies repeated every three sites. We choose this regular pattern of frequencies for simplicity, since each qubit needs to be off-resonant from its neighbours and three subgroups of qubits need to be addressed separately with microwave drives for the Trotterization. The repetition of the same three frequencies is however not needed. A periodic 1D qubit chain can be obtained experimentally, e.g., as a subsystem of IBM Quantum's 2D architectures~\cite{Wang2018}. Extending the parent Hamiltonians of Sec.~\ref{sec:parent Hamiltonians} to open boundary conditions will yield some modification to the annihilation operators at the boundary, but away from the boundary $h_i$ does not change. 

\begin{figure}[t]
\centering
\includegraphics[width=0.7\linewidth]{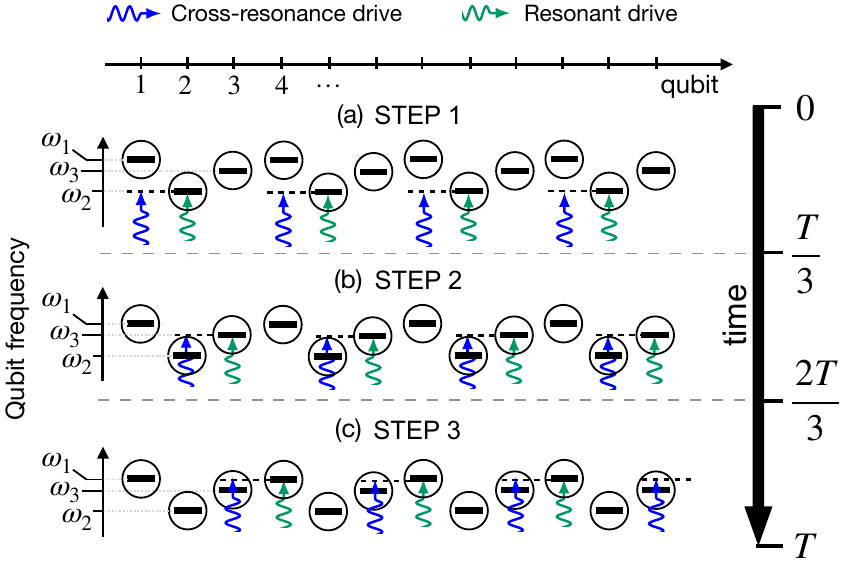}
\caption{Sketch of the basic Lie-Suzuki-Trotter sequence for implementing the parent Hamiltonian of the $x$-polarized state. The sequence involves three sub-steps, in which every third qubit is driven (blue) to activate the cross-resonance interaction with its neighbor to the right. Additional resonant drives (green) compensate for undesired single-qubit terms arising from the cross-resonance effect. Here we only considered a set of three different qubit transition frequencies which repeat along the qubit chain, for simplicity, but the protocol applies equally for the realistic case of fully different frequencies (with a corresponding adjustment of the driving frequencies).}
\label{fig:sketch}
\end{figure}

For the $x$-polarized state (cluster state) the protocol consists of three sub-steps, labelled with $n=1,2,3$, respectively, and sketched in Fig.~\ref{fig:sketch}.
In the $n$th step, every third qubit starting from site $n$ is driven at the frequency of its neighbor to the right (left), along with the single-qubit modifications discussed in Sec.~\ref{subsec: cross resonance effect}, for a duration $T/3$, yielding the Hamiltonian
			\begin{gather} \label{eq:mathcalH1}
				\mathcal{H}_n=\sum_{i=0}^{L/3-1}H_{3i+n},
			\end{gather}
   where $H_i$ is the corresponding modified cross-resonance Hamiltonian of Eq.~\eqref{eq:two site Hamiltonian x-polarized} [Eq.~\eqref{eq:three site Hamiltonian cluster}].
   
Iterating these three steps in a first-order Lie-Suzuki-Trotter sequence will yield a time evolution that is periodic with period $T$. The time evolution operator for a time $T$ is then
	\begin{gather}\label{eq:time evolution}
		U(T)=e^{-i\mathcal{H}_{3}T/3}e^{-i \mathcal{H}_{2}T/3}e^{-i \mathcal{H}_{1}T/3}.
	\end{gather}
Using Baker-Campbell-Hausdorff formulae, we can identify the Floquet Hamiltonian generating the stroboscopic evolution~\cite{Goldman2014,Eckardt2015,Bukov2015}, $U(T)=\exp\left(-iH_{F}T\right)$, as
	\begin{equation}
		H_{F}=\frac{1}{3}(\mathcal{H}_3+ \mathcal{H}_2+\mathcal{H}_1)+ \mathcal{O}(T), \label{eq:effective Hamiltonian}
	\end{equation}
 where $\mathcal{O}(T)$ contains nested commutators of the $\mathcal{H}_i$. For small Trotter step $T$, this Floquet Hamiltonian should have either the $x$-polarized state or the cluster state as an approximate zero energy eigenstate, depending on which modified cross-resonance Hamiltonian is used [Eq.~\eqref{eq:two site Hamiltonian x-polarized} or \eqref{eq:three site Hamiltonian cluster}]. The $x$-polarized and cluster states will not be exact scar states of $H_F$ due to the $\mathcal{O}(T)$ commutator terms and the $\sigma_i^z\sigma_{i+1}^z$ term in the cross-resonance Hamiltonians. These terms should, however, be small, so that the $x$-polarized and cluster states should be close to exact scar states. 
 
\begin{figure}
	\centering
		\includegraphics[width=0.9\linewidth]{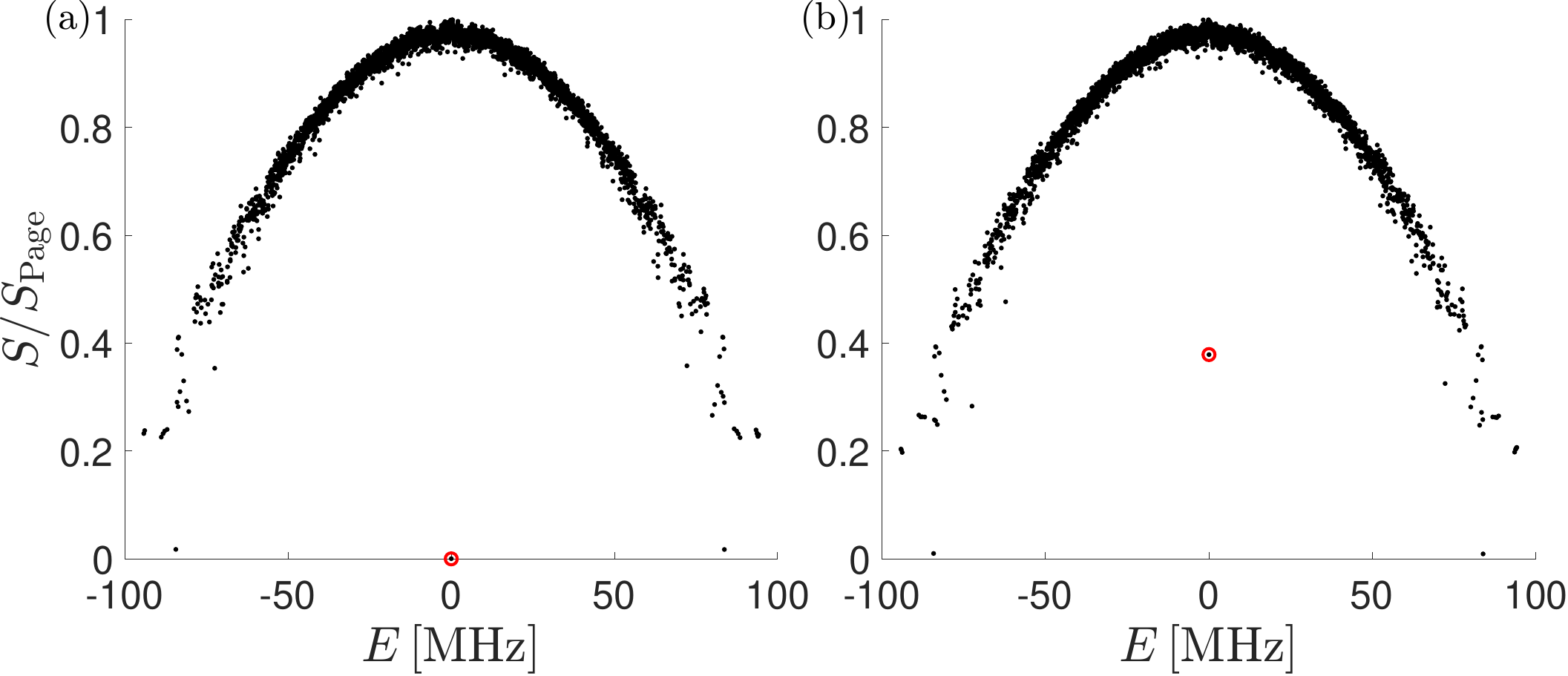}
		\caption{Half-chain von Neumann entanglement entropy divided by the Page entropy as a function of energy for each Floquet mode of the time evolution operator~\eqref{eq:time evolution}, using the modified cross-resonance Hamiltonian (a) Eq.\ \eqref{eq:two site Hamiltonian x-polarized} or (b) Eq.\ \eqref{eq:three site Hamiltonian cluster}. The respective scar state's energy and entropy is indicated with a red circle. Both scar states are quite close in energy and entropy to an exact energy eigenstate but do not match exactly.}
		\label{fig:spectrum}
	\end{figure}

 To confirm this, we report in Fig.\ \ref{fig:spectrum} the entanglement entropy of the numerically-exact Floquet modes as a function of their quasienergy, for both choices of cross-resonance Hamiltonians. The parameter values used are in the regime of small $T$ (namely, $2\pi/T\gg \Omega^2/\Delta_{i,j}$), and they are specified below. The Floquet modes and quasienergies are computed by diagonalizing the end-of-period time evolution operator $U(T)$ of Eq.~\eqref{eq:time evolution}. The red circles in Fig.~\ref{fig:spectrum} indicate the quasienergy and half-chain entanglement entropy of the scar states. The quasienergy spectrum is unwound by selecting, for each Floquet mode, the Floquet copy which is closer to the expectation value of the time-averaged Hamiltonian $(\mathcal{H}_1+\mathcal{H}_2+\mathcal{H}_3)/3$ [that is, the leading term of $H_F$ in Eq.~\eqref{eq:effective Hamiltonian}]. For both the $x$-polarized and the cluster case, one Floquet mode is very close to the scar state, having a probability overlap of $0.9993$ and $0.9992$ for the $x$-polarized state and the cluster state, respectively. 

The parameter values used for this calculation are the same as in Refs.~\cite{PhysRevA.93.060302,PhysRevA.101.052308,Malekakhlagh2020}
	\begin{gather}\nonumber
		\frac{J}{2\pi}=3.8\,\text{MHz},\quad \frac{\Omega}{2\pi}=50\,\text{MHz},\quad \frac{\alpha}{2\pi}=330\,\text{MHz},\\ \label{eq:parameters}	
		\frac{\omega_1}{2\pi}=5.114\,\text{GHz},\quad \frac{\omega_2}{2\pi}=4.914\,\text{GHz},\quad \frac{\omega_3}{2\pi}=5.014\,\text{GHz},	
	\end{gather}
where $\omega_{1,2,3}$ denote the three transmon frequencies which are repeated every three sites. For the $x$-polarized state the order of the transmon frequencies is $[\omega_1,\omega_2,\omega_3,\dots]$ and for the cluster state the order is $[\omega_3,\omega_2,\omega_1,\dots]$. The order of the frequencies is chosen to obtain approximately homogeneous values of the coefficients given in Eqs.~\eqref{eq:coefficients} throughout the spin chain, while using the same driving amplitude $\Omega$ on all sites. Spatially homogeneous values rule out potential disorder-induced effects preventing the thermalization of non-scar eigenstates. For different choices of frequency patterns, spatially homogeneous coefficients can be obtained by making $\Omega$ site-dependent and adapting it accordingly. For the Trotter step $T$ and the system size $L$, we use 
	\begin{gather}
		T\approx 16\ \text{ns},\quad L=12,
	\end{gather}
 corresponding to a dimensionless time $J T\approx 0.4$.
The above parameter values are also used for the time evolution simulated in Sec.~\ref{sec:optimal}. The values of the coefficients from Eq.~\eqref{eq:coefficients} for the cross-resonance Hamiltonians~\eqref{eq:two site Hamiltonian} and~\eqref{eq:three site Hamiltonian} are given in Appendix \ref{app:C}. 

The results of Fig.~\ref{fig:spectrum} about the entanglement entropy confirm the effectiveness of the Lie-Suzuki-Trotter sequence. At the same time, they already provide a signature that can be detected in experiments. Indeed, the behaviour of the von Neumann entropy reported in Fig.~\ref{fig:spectrum} is reflected also, in particular, in the second R\'enyi entropy, $S^{(2)}(\rho)=-\ln[ \mathrm{tr} (\rho^2)]$~\cite{doi:10.1126/science.aaf6725}. The latter can be measured, without the need of full tomography and additional copies of the system, through randomization methods, as proposed in Refs.~\cite{vanEnk2012, Elben2018, Brydges2019}. The procedure involves applying random single-qubit unitaries and then studying statistical correlations in samples of probability distributions for bit-strings obtained by measuring in a fixed qubit basis. This approach has been implemented in superconducting circuits, e.g., in Ref.~\cite{Satzinger2021} for the detection of topologically ordered states. In Sec.~\ref{sec:optimal}, we will propose and analyse additional, more easily accessible signatures.

Here and in the following, we assume that the $x$-polarized and the cluster states can be initially prepared as the starting point of the scar-detection protocol. Assuming that all transmons are initially cooled down to their ground state, the $x$-polarized state $\ket{x}$ is easily prepared with half-a-Rabi oscillation through a resonant microwave drive on each qubit. Starting from this state, the cluster state is obtained according to Eq.~\eqref{eq:cluster_state}, by applying a controlled-$Z$ gate on each pair of neighbouring qubits. The necessary CZ gates may be performed in parallel on non-overlapping pairs and can be implemented, for fixed-frequency transmons, through the cross-resonance effect or other recently proposed native gates~\cite{Wei2024}. 1D cluster states of variable size have been prepared both with fixed-coupling superconducting qubits~\cite{Wang2018, Gong2019} and in architectures with tuneable couplers~\cite{Cao2023}.

\section{Dynamical signatures}\label{sec:optimal}

Defining properties of a scar state are that (1) it is an energy eigenstate, (2) it has low entanglement entropy, and (3) most states near the same energy have entanglement entropy close to the Page entropy as expected for states showing eigenstate thermalization. The Page entropy is $S_{\text{Page}}=\ln(m)-m/2n$ where $m$ and $n$ are the dimensions of the two subsystems, and $1\ll m\leq n$ is assumed~\cite{PhysRevLett.71.1291}.
The entropy plots in Fig.~\ref{fig:spectrum} indicate that the system described by the Floquet Hamiltonian~\eqref{eq:effective Hamiltonian} possesses scar states that are well approximated by the $x$-polarized and the cluster state, respectively. This theoretically confirms that the effective model obtained via Trotterization is a good surrogate of the ideal MPS construction. While such states can be efficiently prepared and while their (R\'enyi) entanglement entropy can be measured in experiments with the methods discussed in Sec.~\ref{sec:eff_ham} [thus confirming property (2)], a similar experiment alone would not be sufficient to conclusively detect the scar. We thus propose to obtain information about properties (1) and (3) with other and simpler methods, based on time evolution. In particular, we propose to study signatures in the quench dynamics under the Floquet Hamiltonian in three different types of experiments.

For all protocols, we assume that the approximate scar (either the $x$-polarized state or the cluster state) is initially prepared, following the procedure discussed at the end of Sec.~\ref{sec:eff_ham}. The Trotter sequence for the corresponding parent Hamiltonian is then launched abruptly, such that the state evolves stroboscopically under the action of $H_{F}$. 
In all types of proposed experiments, the evolution of the scar is benchmarked against the evolution of a local deformation of itself, obtained by applying a unitary rotation to one single qubit. A local deformation of the scar state will generally not change the energy significantly, but the deformed state will have support on many energy eigenstates near the scar state energy and therefore thermalize quickly if the non-scar spectrum is thermal. In one proposed experiment, the evolution is governed simply by $H_{F}$ and local observables are monitored as a function of time, in order to probe their persistence for the scar state as compared to the deformed state. In the second type of experiment, the digital sequence is slightly modified such that the coefficients of the Floquet Hamiltonian contain randomized offsets of controlled maximal strength. Samples of evolutions at different noise level are then used to contrast the sensitivity to noise of the scar in local observables with the insensitivity of its deformed version, in a zero-noise-extrapolation-like analysis. In the third experiment type, the comparison between the scar state and its deformed version is extended to the dependence of the observables on the resolution of the Trotterization. As discussed in the following (Sec.~\ref{sec:resolution}), the scar state is expected to be unusually stable with respect to variations in the length of the Trotter step, differently from generic states.

	\begin{figure}
	\centering
		\includegraphics[width=\linewidth]{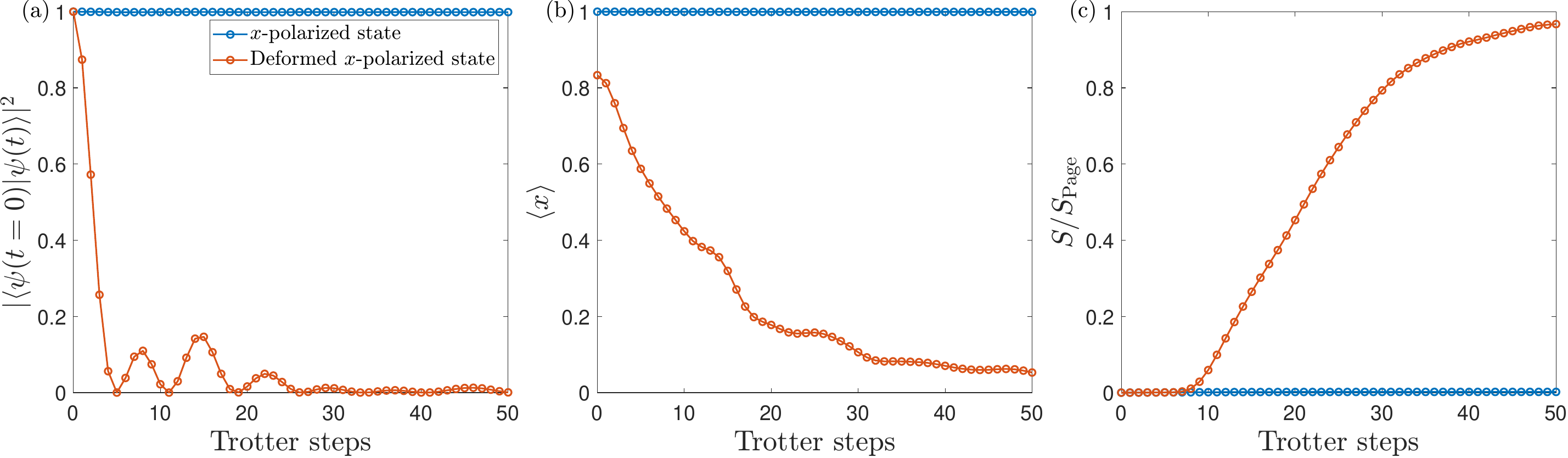}
		\caption{Time evolution of the $x$-polarized state and its deformed version under the Floquet Hamiltonian~\eqref{eq:effective Hamiltonian} with modified cross-resonance Hamiltonian~\eqref{eq:two site Hamiltonian x-polarized}. At each time step, we present (a) the fidelity of the propagated state with respect to the initial state, $|\langle\psi(t=0)|\psi(t)\rangle|^2$, (b) the expectation value of the local operator $x=\frac{1}{L}\sum_{i=1}^L\sigma_i^x$, and (c) the half-chain von Neumann entanglement entropy divided by the Page entropy. The $x$-polarized state (blue) almost does not change under the time evolution, while the deformed state (orange) quickly thermalizes.}
		\label{fig:optimal_x}
	\end{figure}

\subsection{Probing eigenstate properties}\label{sec:eigenstate properties}
In the first experiment type, we propose to compare the time evolution of the scar state with the time evolution of a locally deformed version of itself in terms of two observables. The first is the entanglement entropy. The deformed scar state is expected to thermalize quickly, reaching the Page entropy, while the scar state will remain at low entropy, or at least increase its entropy slowly. The other measurement probes the expectation value of a local operator for which the scar state has a known non-zero value. A scar state, being an energy eigenstate, should preserve the expectation value during the dynamics, while this is not the case for the deformed scar state. Note that the entropy and expectation value for the deformed scar state before the time evolution may in general differ from that of the scar state. 

For both the $x$-polarized and cluster states we choose the deformation to be a $\sigma^y$ unitary applied to the first site of the chain. For both scar states the deformation does not change the energy or the entropy. The local operator we choose for the $x$-polarized state is 
	\begin{gather}\label{eq:x}
		x=\frac{1}{L}\sum_{i=1}^L\sigma_i^x,\quad \langle x\rangle_{x\text{-polarized}}=1.
	\end{gather}
This quantity can be measured straightforwardly through dispersive readout~\cite{krantz2019quantum} after applying a qubit rotation.
For the cluster state, we monitor the local stabilizer operator
	\begin{gather}\label{eq:zxz}
		zxz=\frac{1}{L}\sum_{i=1}^L\sigma_i^z\sigma_{i+1}^x\sigma_{i+2}^z,\quad \langle zxz\rangle_{\text{cluster}}=-1.
	\end{gather}
    	\begin{figure}
	\centering
		\includegraphics[width=\linewidth]{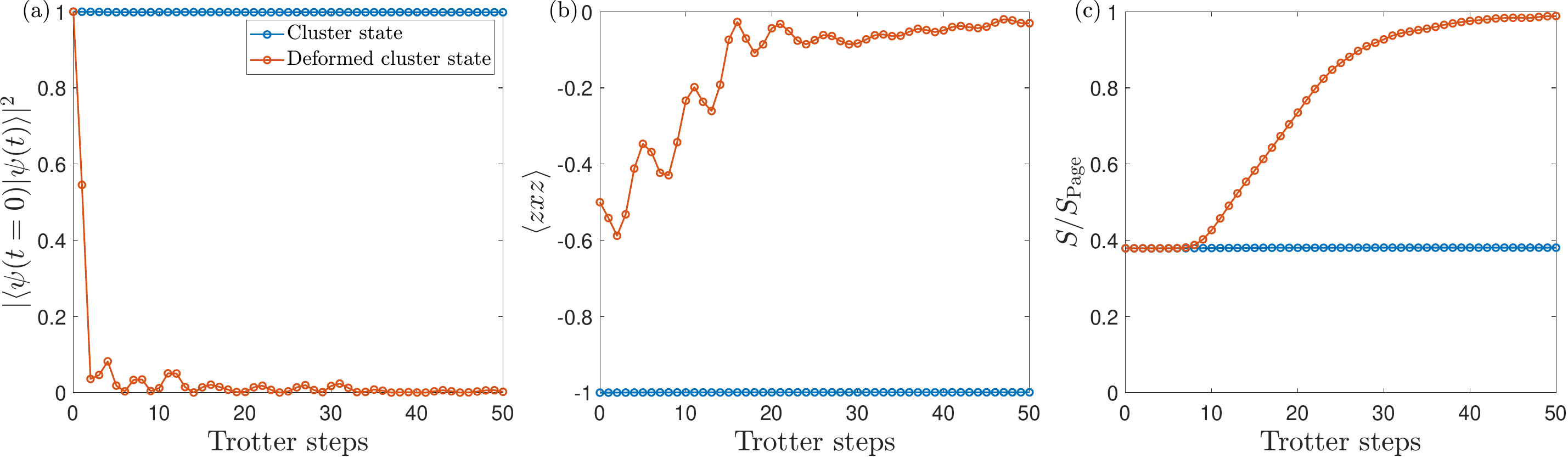}
		\caption{Time evolution of the cluster state and its deformed version under the Floquet Hamiltonian~\eqref{eq:effective Hamiltonian} with modified cross-resonance Hamiltonian~\eqref{eq:three site Hamiltonian cluster}. At each time step, we present (a) the fidelity $|\langle\psi(t=0)|\psi(t)\rangle|^2$, (b) the expectation value of the local operator $zxz=\frac{1}{L}\sum_{i=1}^L\sigma_i^z\sigma_{i+1}^x\sigma_{i+2}^z$, and (c) the half-chain von Neumann entanglement entropy divided by the Page entropy. The cluster state (blue) almost does not change under the time evolution, while the deformed state (orange) quickly thermalizes.}
		\label{fig:optimal_cluster}
	\end{figure}
The three-qubit operators can be measured by performing suitable single-qubit rotations and controlled-$Z$ gates before a joint qubit readout~\cite{Gong2019}.
In addition to the local operator expectation values, we also compute the fidelity $|\langle\psi(t=0)|\psi(t)\rangle|^2$ between each state at time $t=0$ and time $t$. This is a direct measure of how much the states change, but it cannot be easily measured in experiments. Here we can use it as an indicator in how far the conservation of the expectation value of the local operator indeed indicates a conservation of the state, for the models considered here. In the actual experiment, measurements need only be taken at the end of the time evolution, but in Fig.~\ref{fig:optimal_x} and \ref{fig:optimal_cluster} we show the fidelity, local operator expectation value, and entanglement entropy as a function of time to have the full picture. As expected based on the entropy plots of Fig.~\ref{fig:spectrum}, both the $x$-polarized state and cluster state are to a very good approximation scar states of their respective Floquet Hamiltonians. Both states almost do not change under the time evolution, while their deformed versions quickly thermalize and reach the Page entropy around $t=30T$. The fidelity of the deformed states approaches zero faster than the local operator expectation values, but after some time (around $t=20T$ to $t=30T$) both are close to zero. Therefore, as long as measurements are taken after an appropriate time, the expectation values well represent the actual change of the deformed states.

\subsection{Sensitivity to protocol distortion}
\label{sec:error}

	\begin{figure}
	\centering
		\includegraphics[width=\linewidth]{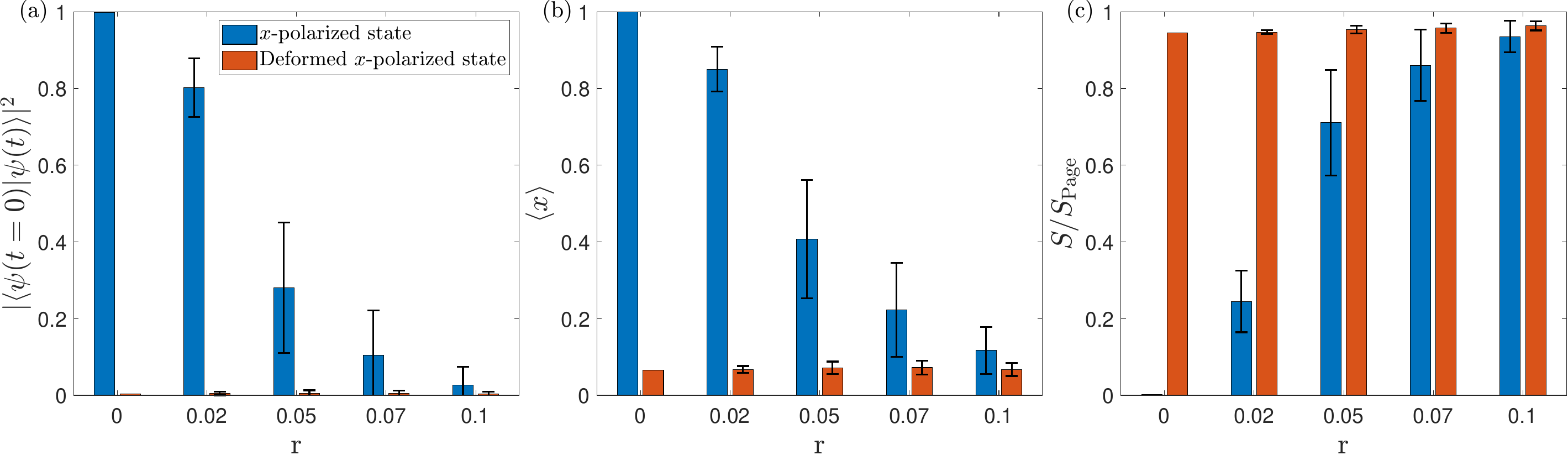}
		\caption{Time evolution of the $x$-polarized state and its deformed version driven by the Floquet Hamiltonian, with random errors of varying error strength $r$. The time evolution is run 500 times for each error strength (excluding $r=0$) and the results, sampled at time $t=30T$ are averaged, with error bars indicating the standard deviation. The quantities calculated are (a) the fidelity $|\langle\psi(t=0)|\psi(t)\rangle|^2$, (b) the expectation value of the local operator $x=\frac{1}{L}\sum_{i=1}^L\sigma_i^x$, and (c) the half-chain von Neumann entanglement entropy divided by the Page entropy. With increasing error strength the $x$-polarized state starts to decay, however for errors below about $0.05$, it is still noticeably more stable than the deformed state.}
		\label{fig:error_x}
	\end{figure}

The time evolution of the $x$-polarized and cluster states under the Floquet Hamiltonian~\eqref{eq:effective Hamiltonian} yields a measurable difference between the scar states and their deformed versions, which could be used to experimentally observe the scars. However, experimental imperfections may yield additional errors, not taken into account in the model used above. The modified cross-resonance Hamiltonians of Eqs.\ \eqref{eq:two site Hamiltonian x-polarized} and \eqref{eq:three site Hamiltonian cluster} only reproduce the annihilation operators (annihilating the scar states), when the different coefficients $c_{i,j}^\alpha$ take the correct values with respect to one another, imposed by Eqs.~\eqref{eq:x-polarized} and \eqref{eq:cluster}. Potential errors in a real experiment may offset this balance leading to the scar states thermalizing and ultimately not being distinguishable from their deformed versions. 
It is thus important to quantify the impact of errors on the observation of the scar states. From a different perspective, if the experimental noise is not strong enough to fully wash out any signature of scarring, one may add noise artificially to perform a zero-noise-extrapolation-type experiment~\cite{Kandala2019}. Namely, one introduces additional \textit{controlled noise} with a variable maximal magnitude and monitors how the desired signatures progressively degrade as the error strength is increased. The ideal value may then be estimated by extrapolating to the zero-noise limit. 

With these motivations, we next study the time evolution with distorted driving protocols. We do so by including randomized offsets in the values of the coefficients of the modified cross-resonance Hamiltonians of Eqs.~\eqref{eq:two site Hamiltonian x-polarized} and \eqref{eq:three site Hamiltonian cluster}. The coefficients, both for the modifications and the cross-resonance effect, are changed according to
	\begin{gather}
		c_{i,j}^\alpha \rightarrow (1+r u_{i,j}^\alpha)c_{i,j}^\alpha,
	\end{gather}
where $r$ is the maximal strength of the error, $u_{i,j}^\alpha$ are uniformly distributed random numbers in the range $-1\le u_{i,j}^\alpha\le 1$, and $\alpha$ denotes $\alpha \in \{z, x,zx, zz\}$. This type of error does not necessarily mimic actual experimental errors in detail, but serves as a guideline for how robust the observation of the scar states is to errors in general. Similar protocol distortions may be deliberately induced by altering the noise-free value of the driving amplitude or frequency of the cross-resonance and resonant drives.

	\begin{figure}
		\includegraphics[width=\linewidth]{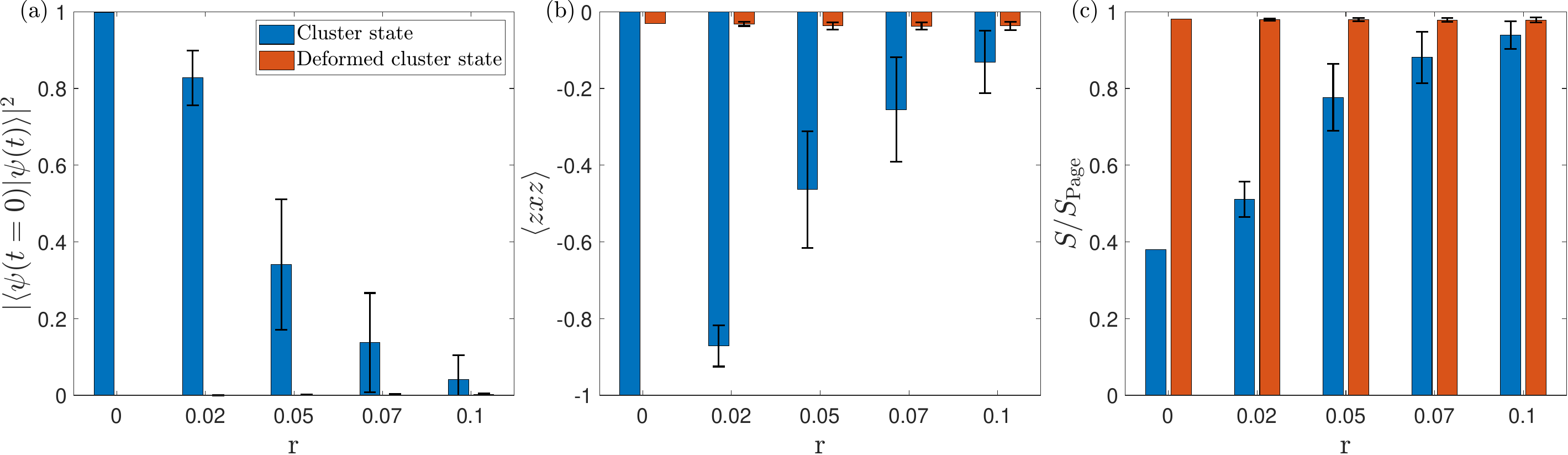}
		\caption{Time evolution of the cluster state and its deformed version under the Floquet Hamiltonian of Eq.\ \eqref{eq:effective Hamiltonian} with random errors of varying maximal strength $r$. The time evolution is run 500 times for each error strength (excluding $r=0$) and the results, sampled at time $t=40T$ are averaged, with error bars indicating the standard deviation. The quantities calculated are (a) the fidelity $|\langle\psi(t=0)|\psi(t)\rangle|^2$, (b) the expectation value of the local operator $zxz=\frac{1}{L}\sum_{i=1}^L\sigma_i^z\sigma_{i+1}^x\sigma_{i+2}^z$, and (c) the half-chain von Neumann entanglement entropy divided by the Page entropy. With increasing error strength the cluster state starts to decay, however for errors below about $0.05$, it is still noticeably more stable than the deformed state.}
		\label{fig:error_cluster}
	\end{figure}

In the simulations presented below, we run the time evolution like in Sec.\ \ref{sec:eigenstate properties}, sampling the fidelity, local operator expectation value, and entropy after a time $t=40T$. For this simulation, $H_{F}$ is approximated with the analytical estimate of Eq.~\eqref{eq:effective Hamiltonian} up to first order in $T$ included, to reduce computation time. The $x$-polarized and cluster states are still approximate scar states of this lower order effective Hamiltonian having probability overlaps of $0.9992$ and $0.9962$, respectively, with a highly excited low entropy Floquet mode. The time evolution is repeated for different random numbers $u_{i,j}^\alpha$ and the results are averaged. 
The average results for $r=0,0.02,0.05,0.07,0.1$ are shown in Fig.\ \ref{fig:error_x} and \ref{fig:error_cluster} for the $x$-polarized state and cluster state, respectively. The error bars show the standard deviation. The deformed states generally behave similarly to one another, exhibiting similar averages and standard deviations. The latter are small compared to those of the approximate scar states, as expected for thermal states. The sensitivity of the scar states is evident in the decay as a result of the errors, with larger error strength leading to a stronger decay. Moreover, the fact that the standard deviations of the approximate scar states are noticeably larger compared to those of the deformed states indicates that these states are also sensitive to the details of the errors. However, roughly for error strengths $r$ up to 0.05, they still thermalize noticeably slower than the deformed states. If the random number $u_{i,j}^\alpha$ are such that the cross-resonance Hamiltonians are still close to the annihilation operators (just rescaled), then the $x$-polarized and cluster states are still approximate scar states. The different response of the scar and deformed states to induced errors thus represents a valuable signature that can be probed in experiments to reveal the scarring phenomenon.

\subsection{Sensitivity to coarse Trotterization}
\label{sec:resolution}

\begin{figure}
    \centering
    \includegraphics[width=\linewidth]{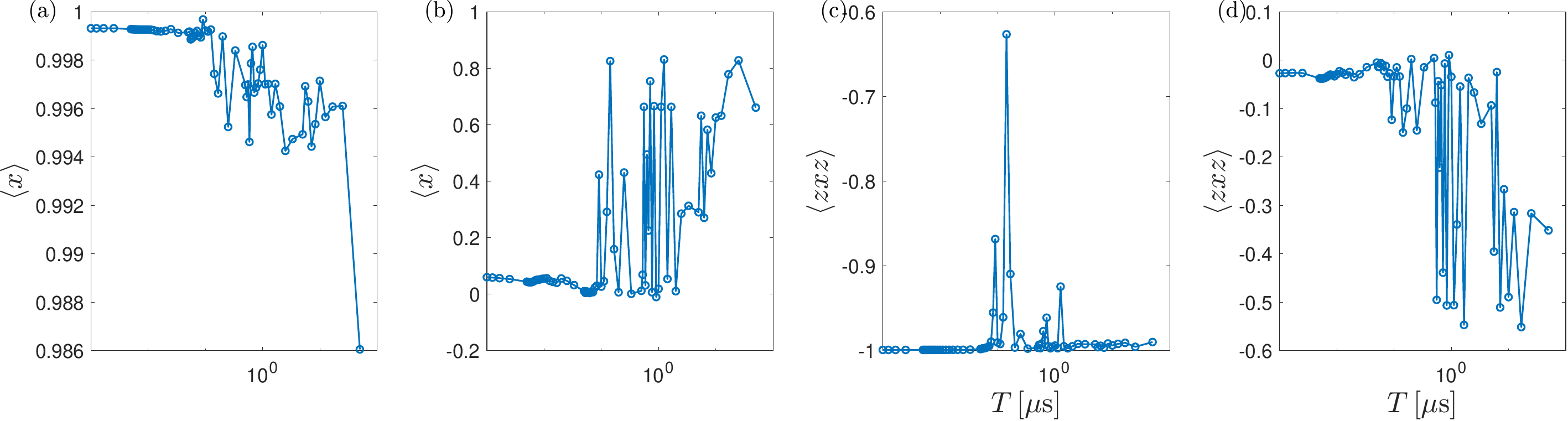}
    \caption{Plot of the expectation value of the local observable for (a) the $x$-polarized state, (b) the deformed $x$-polarized state, (c) the cluster state, and (d) the deformed cluster state [Eq.~\eqref{eq:x} and~\eqref{eq:zxz} respectively] plotted against the size of the Trotter step. In all cases the expectation values are sampled after time evolving the state for a time $t=5\ \mu\text{s}$. The scar states are generally less sensitive to the size of the Trotter step (especially the $x$-polarized state) while the deformed states show a sudden change in expectation value around $T=10^{-1}\ \mu\text{s}$.}
    \label{fig:trotter step}
\end{figure}

The final comparison involves probing the sensitivity of local observables, for the scar state and its deformed partner, with respect to the resolution of the Trotterization, namely the length $T$ of the Trotter step. As $T$ is increased, the high-frequency description of Eq.~\eqref{eq:effective Hamiltonian} is expected to break down. The stroboscopic dynamics is, then, not well captured by considering only the leading terms of the Baker-Campbell-Hausdorff expansion for the Floquet Hamiltonian. Hence, the expectation values of the local observables for generic initial states should quickly drift away from the value predicted using the time-averaged Hamiltonian $(\mathcal{H}_1+\mathcal{H}_2+\mathcal{H}_3)/3$. 

However, a different behaviour can be expected for the scar state. Indeed, recall that the scarred model is obtained by constructing operators $h_i$ which all annihilate the scar, Eq.~\eqref{eq:sumhi}. Meanwhile, the Lie-Suzuki-Trotter decompositions proposed in Sec.~\ref{sec:eff_ham} involves stepwise Hamiltonians $\mathcal{H}_n$ which are sums of operators approximating $h_i$ on disjoint subsystems [see Eqs.\ \eqref{eq:mathcalH1}, \eqref{eq:two site Hamiltonian x-polarized} and \eqref{eq:three site Hamiltonian cluster}]. It follows that the higher-order terms in $T$ in the Floquet Hamiltonian \eqref{eq:effective Hamiltonian}, given by the Baker-Campbell-Hausdorff expansion, approximately involve nested commutators of the $h_i$ of the form $[h_i, \dots ,[h_j,h_k]\dots]]$. Since the $h_i$ annihilate the scar state, so do also all such nested commutators. Therefore, one can expect that, even at large $T$, the expectation values for the scar state at stroboscopic times should be weakly affected by the coarseness of the Trotterization.

This behaviour is indeed confirmed in simulations, as reported in Fig.~\ref{fig:trotter step} for both the $x$-polarized and the cluster model. Even well beyond the radius of convergence of the small-$T$ expansion, the expectation values do not vary by much for the scar state [Fig.~\ref{fig:trotter step}(a) and (c)]. On the contrary, they quickly depart from the small-$T$ effective description for the deformed state [Fig.~\ref{fig:trotter step}(b) and (d)]. Probing this effect in an experiment would provide an additional dynamical signature of single-state scarring.

\section{Conclusion}\label{sec:conclusion}

We have proposed a blueprint for an experimental implementation of two models with a single scar state each and protocols to experimentally observe them with fixed-frequency, fixed-coupling superconducting qubits. In these models, the scar states are the $x$-polarized state and the 1D cluster state, respectively. The experimental implementation is based on Lie-Suzuki-Trotter sequences of modified cross-resonance drives of superconducting qubits. The resulting Floquet Hamiltonians computed numerically have been used to simulate the stroboscopic time evolution of the scar states and of a locally deformed version of them. Using entanglement entropy and the expectation value of local observables, the scar states are shown to thermalize slowly while the deformed states thermalize quickly. Probing this behaviour in experiments would provide strong evidence of single-scar-state physics. 

An interesting outlook is the extension of the protocols proposed here for observing a scar-state phase transition. This is a transition in which the scar state always remains an energy eigenstate in the middle of the spectrum with low entanglement, but its properties undergo a phase transition as a parameter changes~\cite{PhysRevResearch.6.L042007}. The model described in appendix~\ref{app:A} features a scar-state phase transition at $g=0$, where the annihilation operators $h_i$ in the Hamiltonian of Eq.~\eqref{eq:sumhi} take the form 
\begin{equation} \label{eq:ghzH}
h_i = - 2\sigma_i^z + \sigma_{i+1}^z +\sigma_i^z \sigma_{i+1}^x - \sigma_{i+1}^x\sigma_{i+2}^z +\sigma_i^z \sigma_{i+1}^z \sigma_{i+2}^z,
\end{equation}
and the scar is a GHZ state~\cite{PhysRevResearch.6.L042007}. 
The key challenges toward observing this phenomenon are twofold. They regard how to realize the Hamiltonian~\eqref{eq:ghzH} around $g=0$ and how to drive the system along the instantaneous single scar state without transitions to thermal states. Concerning the Hamiltonian, the challenge is realizing the combination of two- and three-qubit couplings in Eq.~\eqref{eq:ghzH}. Two possible strategies to address this problem are (i) to use higher-order Trotterization to decompose the three-qubit interaction into a sequence of cross-resonance gates and single-qubit unitaries or (ii) the use of optimized Floquet engineering methods proposed in Ref.~\cite{Petiziol2021, Petiziol2024tc}, which, compared to Trotterization, may allow for the achievement of stronger effective three-body terms and fewer pulsing steps.

Concerning how to drive the scar state transitionlessly, the problem is that the scar is not a ground state protected by a gap, but an excited state surrounded by thermal states. This strongly challenges the application of adiabatic methods. However, a recent study has revealed that an adiabatic sweep is actually possible, thanks to the very properties of scar states, and that it can be robust against small perturbations at moderate system sizes~\cite{PRXQuantum.5.020365}. An adiabatic variation of the Lie-Suzuki-Trotter sequence considered here, extended to include the $g=0$ Hamiltonian, may then be a promising approach toward realizing the single-scar-state phase transition.

\section*{Acknowledgements}


\paragraph{Funding information}
This work has been supported by Carlsbergfondet under Grant No.\ CF20-0658 and by the Deutsche Forschungsgemeinschaft (DFG, German Research Foundation) via the Research Units FOR 2414 (projectNo.~277974659) and FOR 5688 (project No.~521530974). F.~P. acknowledges funding from the Deutsche Forschungsgemeinschaft (DFG, German Research Foundation) through the Emmy Noether Programme -- project number 555842149. The authors thank the anonymous referee 2 for pointing out the connection between single quantum many-body scars and Hellers' original concept of quantum scars as well as Roland Ketzmerick for  helpful comments on this issue.

\paragraph{Code availability} The code used for the numerical simulations is available from the authors upon reasonable request.

\begin{appendix}
\numberwithin{equation}{section}

\section{General form of the annihilation operators of the single-scar model}
\label{app:A}
In this appendix we review the general form of the annihilation operators $h_i$ discussed in Sec.~\ref{sec:parent Hamiltonians} which are given by~\cite{PhysRevResearch.6.L042007}\begin{gather}\label{eq:annihilation operator}h_i=\sum_{n=1}^4(-1)^n|\psi_n\rangle\langle\psi_n|,
	\end{gather}
where the local states $|\psi_n\rangle$, before normalization, are defined by
	\begin{subequations}
		\begin{align}
		 |\psi_1\rangle\propto & -g|{{\uparrow}}{{\uparrow}}{{\uparrow}}\rangle-|{{\uparrow}}{{\uparrow}}{{\downarrow}}\rangle+|{{\uparrow}}{{\downarrow}}{{\uparrow}}\rangle+|{{\uparrow}}{{\downarrow}}{{\downarrow}}\rangle, \\
			|\psi_2\rangle\propto& -|{{\downarrow}}{{\uparrow}}{{\uparrow}}\rangle-|{{\downarrow}}{{\uparrow}}{{\downarrow}}\rangle+|{{\downarrow}}{{\downarrow}}{{\uparrow}}\rangle+g|{{\downarrow}}{{\downarrow}}{{\downarrow}}\rangle,\\
			|\psi_3\rangle\propto & -ga|{\uparrow}{\uparrow}{\uparrow}\rangle
			+\frac{a}{2}\left(1+g^2\right)|{\uparrow}{\uparrow}{\downarrow}\rangle 
			 +a|{\uparrow}{\downarrow}{\uparrow}\rangle 
			 -\frac{a}{2}\left(1+g^2\right)|{\uparrow}{\downarrow}{\downarrow}\rangle 
			 -\frac{1-a}{2}\left(1+g^2\right)|{\downarrow}{\uparrow}{\uparrow}\rangle \nonumber \\
			& +\left(1-a\right)|{\downarrow}{\uparrow}{\downarrow}\rangle
		 	+\frac{1-a}{2}\left(1+g^2\right)|{\downarrow}{\downarrow}{\uparrow}\rangle 
			 -g\left(1-a\right)|{\downarrow}{\downarrow}{\downarrow}\rangle,\\
			|\psi_4\rangle\propto & -g\left(1-a\right)|{\uparrow}{\uparrow}{\uparrow}\rangle
			+\frac{1-a}{2}\left(1+g^2\right)|{\uparrow}{\uparrow}{\downarrow}\rangle  
			+\left(1-a\right)|{\uparrow}{\downarrow}{\uparrow}\rangle
			-\frac{1-a}{2}\left(1+g^2\right)|{\uparrow}{\downarrow}{\downarrow}\rangle\nonumber \\
    & +\frac{a}{2}\left(1+g^2\right)|{\downarrow}{\uparrow}{\uparrow}\rangle 
			-a|{\downarrow}{\uparrow}{\downarrow}\rangle  
			-\frac{a}{2}\left(1+g^2\right)|{\downarrow}{\downarrow}{\uparrow}\rangle
			+ga|{\downarrow}{\downarrow}{\downarrow}\rangle,
		\end{align}
	\end{subequations}
where $a$ is a free parameter chosen between zero and one. We indicated with $\ket{{\uparrow}}$ and $\ket{{\downarrow}}$ the $+1$ and $-1$ eigenstates of $\sigma_z$, respectively. In this work we choose $a=1$ and $g=\pm 1$, which yields a simple form of $h_i$ and thus of $H$.

\section{Definition of the coefficients $\nu$}\label{app:B}
In this appendix we report the definition of the $\nu$ coefficients used in computing the coefficients of the effective two qubit Hamiltonian of the cross-resonance effect, Eq.~\eqref{eq:coefficients}. As detailed in Ref.~\cite{Malekakhlagh2020}, the $\nu$ coefficients are related to the matrix elements of the charge operator of the transmon in the energy basis. When approximating the transmon as a four level system, there are four relevant $\nu$ coefficients which are given by
\begin{subequations}
    \begin{align}
        \nu_{01}& =1-\frac{1}{8}\epsilon-\frac{11}{256}\epsilon^2+\mathcal{O}\left(\epsilon^3\right),\\
        \nu_{12}& =\left(1-\frac{1}{4}\epsilon-\frac{73}{512}\epsilon^2\right)\sqrt{2}+\mathcal{O}\left(\epsilon^3\right),\\
        \nu_{23} & =\left(1-\frac{3}{8}\epsilon-\frac{79}{256}\epsilon^2\right)\sqrt{3}+\mathcal{O}\left(\epsilon^3\right),\\
        \nu_{03} & =-\frac{\sqrt{6}}{16}\epsilon-\frac{5\sqrt{6}}{128}\epsilon^2+\mathcal{O}\left(\epsilon^3\right),
    \end{align}
\end{subequations}
where $\epsilon$ is the unitless anharmonicity measure, which can be found by solving the quadratic equation
\begin{equation}
    \left[9-4\left(\frac{\alpha}{\omega}\right)\right]\epsilon^2+16\left[1-\left(\frac{\alpha}{\omega}\right)\right]\epsilon+64\left(\frac{\alpha}{\omega}\right)=0,\quad \epsilon>0,
\end{equation}
where $\omega$ is the harmonic frequency and $\alpha$ is the anharmonicity of the transmon. When projecting from the four-level system to the two-level qubit system, only the coefficients $\nu_{01}$ and $\nu_{12}$ survive.

\section{Hamiltonian coefficients}\label{app:C}
In Tab.~\ref{tab:Hamiltonian coeeficients} we list the values of the coefficients in the two-site and three-site Hamiltonians \eqref{eq:two site Hamiltonian} and \eqref{eq:three site Hamiltonian}, given the parameter choices of Eq.~\eqref{eq:parameters}. Three values are listed for each coefficient, corresponding to the three different frequencies used in the Trotterization. The values of the single-qubit corrections are found directly from the coefficients of Tab.~\ref{tab:Hamiltonian coeeficients}.

\begin{table}
\centering
    \caption{The value of the coefficients of Eq.~\eqref{eq:two site Hamiltonian} and \eqref{eq:three site Hamiltonian} computed with the parameters of Eq.~\eqref{eq:parameters}. The different coefficients of the different sites $i$ correspond to the three frequencies used in the Trotterization.}
    \label{tab:Hamiltonian coeeficients}
    \begin{tabular}{l| l l l}
        \hline\hline
          & \ \ $ i\text{ mod }3=1\quad$ & $i\text{ mod }3=2\quad$ & $i\text{ mod }3=3$ \\
         \hline
        & \multicolumn{3}{c}{$x$-polarized}\\
          $c_{i,i+1}^{z}/2\pi$ & \ \ $-14.4$ MHz & $9.21$ MHz & $9.21$ MHz \\
         $c_{i,i+1}^{x}/2\pi$ &\ \  $1.25$ MHz & $0.376$ MHz & $0.377$ MHz \\
         $c_{i,i+1}^{zx}/2\pi$ & \ \ $-8.47$ MHz & $5.43$ MHz & $5.44$ MHz \\
         $c_{i,i+1}^{zz}/2\pi$ & \ \ $0.115$ MHz & $0.080$ MHz & $0.080$ MHz \\
         \hline
       &  \multicolumn{3}{c}{cluster}\\
         $c_{i,i+1}^{z}/2\pi$ & \ \ $-14.4$ MHz & $9.21$ MHz & $9.21$ MHz \\
         $c_{i+1,i}^{z}/2\pi$ & \ \ $-14.4$ MHz & $9.21$ MHz & $9.21$ MHz \\
         $c_{i+1,i}^{x}/2\pi$ & \ \ $1.25$ MHz & $0.377$ MHz & $0.376$ MHz \\
         $c_{i+1,i}^{xz}/2\pi$ & \ \ $-8.47$ MHz & $5.44$ MHz & $5.43$ MHz \\
         $c_{i+1,i}^{zz}/2\pi$ & \ \ $0.115$ MHz & $0.080$ MHz & $0.080$ MHz \\
         \hline\hline
    \end{tabular}
\end{table}

\end{appendix}



\bibliography{bibfile.bib}


\end{document}